# Twist-diameter coupling drives DNA twist changes by salt and temperature

Chen Zhang[1,#], Fujia Tian[2,#], Ying Lu[3], Bing Yuan[4], Zhi-Jie Tan[5], Xing-Hua Zhang[1,*], and Liang Dai[2,*]

[1]*College of Life Sciences, Wuhan University, Wuhan 430072, China*
[2]*Department of Physics, City University of Hong Kong, Hong Kong 999077, China*
[3]*Institute of Physics, Chinese Academy of Sciences, Beijing 100190, China*
[4]*School of Physical Science and Technology, Soochow University, Suzhou 215006, China*
[5]*School of Physics and Technology, Wuhan University, Wuhan 430072, China*

DNA deformations play crucial roles in many biological processes and material applications. During DNA deformation, DNA structural parameters often exhibit non-trivial and counterintuitive couplings, such as the twist-stretch and twist-bending couplings. Here, we reveal an unexpectedly strong negative twist-diameter coupling through the synergy of magnetic-tweezers experiments, atomistic molecular dynamics simulations, and theoretical calculations. In experiments, the DNA twist angle always increases with the concentration of NaCl, KCl, or RbCl. Our simulations quantitatively reproduce salt-induced twist changes and reveal the underlying physical mechanism: the reduction of DNA diameter under a high salt concentration leads to the increase in DNA twist angle through a strong negative twist-diameter coupling. The twist-diameter coupling is mediated by two dihedral angles $\chi$ and $\delta$ in DNA structure and the coupling constant is $4.5 \pm 0.8$ $k_B$T/(deg·nm) for one base-pair. Based on this coupling constant, we predict the temperature-dependence of DNA twist $\Delta\omega_{\mathrm{bp}}/\Delta T \approx -0.0102$ deg/K, which agrees with our and previous experimental results. Our analysis suggests that the twist-diameter coupling is a common driving force for salt- and temperature-induced DNA twist changes.

Biological functions and material applications of DNA depend on the response of DNA structures to many factors, such as force [1-3], temperature [4,5], ionic conditions [6-10], because even subtle structural changes in DNA can lead to great effects in many cases, such as DNA-protein interactions [11,12] and DNA nanoparticles from self-assembly [13,14]. DNA structural changes are usually quantified by the variations in geometric parameters, including the twist angle, bending angle, diameter, and stretch. An intriguing phenomenon is that some of these structural parameters are coupled, such as the twist-stretch coupling [1,15-18] and twist-bend coupling [19,20]. These couplings make significant impacts on biological processes of DNA [11,21], such as DNA packaging in vivo [19].

Despite the established experimental studies of the twist-stretch coupling [1,17] and twist-bend coupling [20], little progress has been made regarding the twist-diameter coupling, while DNA twist and diameter are among the most important DNA structural parameters. The slow progress could be attributed to that DNA diameter is difficult to be directly measured in single-DNA experiments.

In this work, we aim to reveal the twist-diameter coupling through a synergic approach of single-DNA magnetic-tweezers (MT) experiments, all-atom molecular dynamics (MD) simulations, and theoretical calculations. In MT experiments, we measured the change in DNA twist angle upon increasing the salt concentration. Despite that DNA diameter cannot be directly measured by MT, it is expected that DNA diameter changes with the salt concentration because DNA diameter is mainly determined by the distance between the negatively charged phosphate groups on two strands, while the electrostatic interactions between phosphate groups can be effectively tuned by the salt concentration. With MD simulations and theoretical calculations, we reproduced the experimental results and analyzed the underlying physical mechanisms.

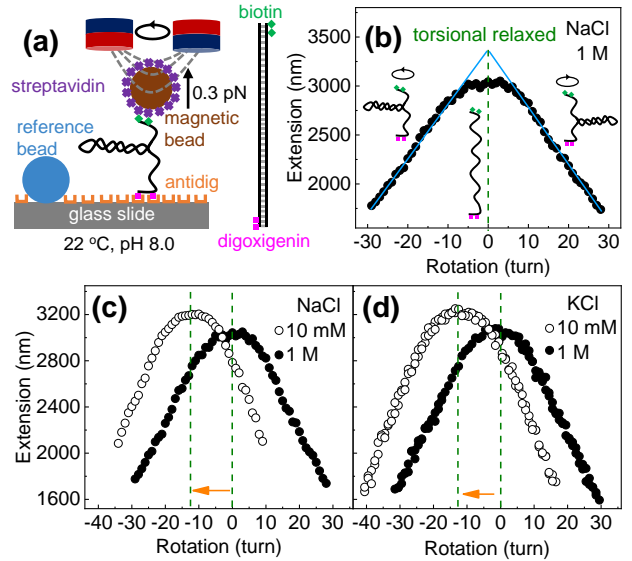

FIG.1 Measuring the changes in DNA twist caused by the increase in salt concentration by MT. (a) MT setup. One end of the DNA is labeled with multiple digoxigenin groups and anchored to a glass slide. The other end of the DNA is labeled with multiple biotin groups and attached to a super-paramagnetic microbead. A pair of NdFeB magnets is used to rotate DNA at a constant force of 0.3 pN. (b) Determination of the torsionally relaxed point of DNA by the symmetrical centerline (dashed green line) of the rotation-extension curve. The two linear fits of the plectoneme range (cyan lines) meet at the torsionally relaxed point. (c-d) Determination of the change in DNA twist calculated using the shift in torsionally relaxed point caused by the decrease in salt concentration.

We first measured the changes in DNA twist caused by increasing the salt concentration using MT [4,6] as illustrated in Fig 1(a). We performed all the experiments at 22 °C, 10 mM Tris-HCl pH 8.0 and 0.3 pN for different salt concentrations (see more details in Sec. S1 of the SI). We rotated a single torsion-constrained double-stranded DNA by magnetic fields, measured the DNA extension simultaneously, and eventually obtained the rotation-extension curves as shown in Fig 1(b). For each rotation-extension curve, we determined the torsionally relaxed point of DNA (dashed green line) by the crossing point of the two linear fits of the plectoneme range (solid cyan lines in Fig 1(b)). The number of rotation-turns of the torsionally relaxed DNA is denoted as $N^*_{turn}$, which changes with the salt concentration (dashed green lines in Fig 1(c)-(d)). Accordingly, we calculated the equilibrium DNA twist angle per base pair, $\omega_{exp}$, from $N^*_{turn}$ through $\omega = (N^*_{turn} \times 360^o)/N_{bp}$, where $N_{bp} \approx 13.6 \times 10^3$ is the number of base pairs in the DNA. Because of the large $N_{bp}$, we can determine $\Delta\omega$ with the resolution on the order of 0.01 deg/bp. Note that our experiment only gives the relative twist angle, i.e., DNA twist change induced by the variation of the salt concentration, $c_{salt}$.

Fig 2(a) shows the experimental results of the relative DNA twist angle, $\Delta\omega$, as a function of $c_{salt}$ for three salts: NaCl, KCl and RbCl. We set $\Delta\omega = 0$ at $c_{salt} = 1$ mol/L so that the data in Fig 2(a) correspond to the change in DNA twist comparing to the value at 1 mol/L. We find that $\Delta\omega$ exhibits similar behavior for all three salts and agrees with the previous results based on DNA supercoiling measurements [22]. These results suggest that the salt-induced twist change should be largely caused by the electrostatic screening effect of ions, while the subtle differences in the salt-induced twist changes among these three salts may be caused by the small differences in ion binding or ion distribution around DNA [7,23-26].

To reveal the molecular mechanism for the salt-induced twist change in our experiments, we performed all-atom MD simulations of 25-bp B-form DNA with the sequence [27]: CGACTCTACGGAAGGGCATCTGCGC. The simulations were implemented in the GROMACS 2018.4 software [28] using OL15 force field [29] (Sec. S2). Fig 3(a) presents one DNA structure from our simulation snapshot. For each concentration of NaCl, KCl or RbCl, we performed 600 ns simulation and calculated DNA twist angles using the program Curves+ [30]. As shown in Fig 2(b-d), our simulations quantitatively reproduced the experimental salt-induced twist changes. We did not simulate $c_{salt} = 0.01$ mol/L because such low concentration requires a large simulation box, which is computationally too expensive.

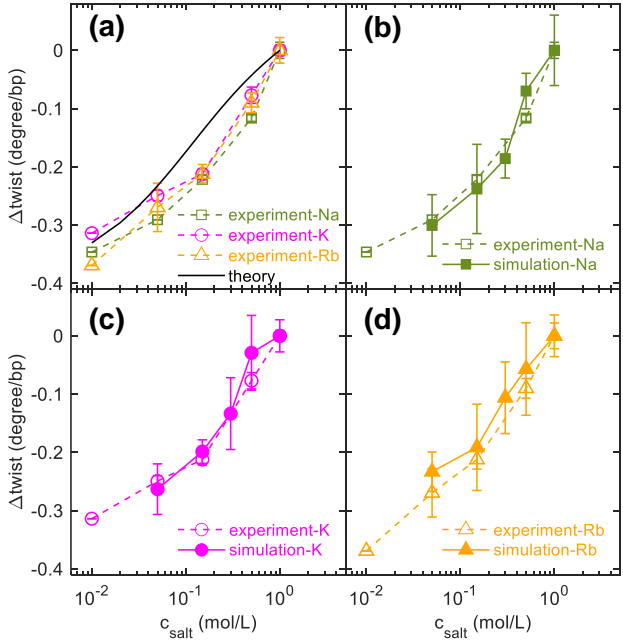

FIG.2 Twist angle as a function of the salt concentration determined by experiments, simulations, and theoretical calculations. The standard deviations obtained from more than three independent experiments or five simulation trajectories of 100 ns are shown in error bars. (a) Experimental results of the twist angle as a function of the salt concentration for NaCl, KCl and RbCl. The black line is the theoretical result from Eq (5). (b-d) Comparison of experimental results and simulation results for three salt species.

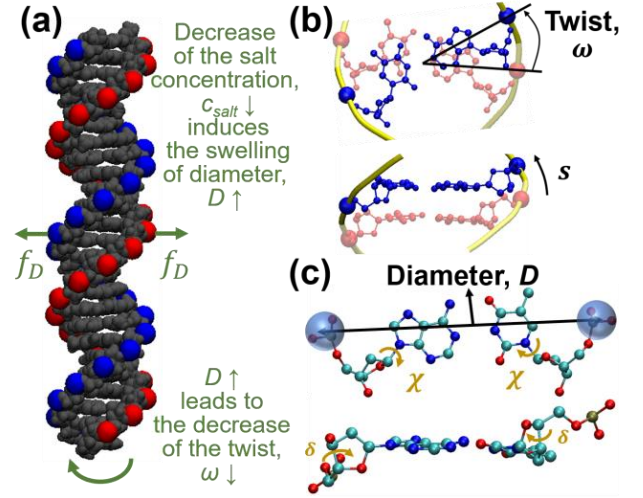

FIG.3 The mechanism underlying the twist-diameter coupling. (a) Illustration of the salt-induced twist change. The decrease of $c_{salt}$ induces the swelling of DNA diameter, which leads to the decrease of DNA twist through the negative twist-diameter coupling. The DNA structure is from our simulation snapshot. The red and blue beads represent negatively charged phosphate groups on two strands, which repel each other through electrostatic repulsions. (b) Top view and side view of one base-pair step to illustrate the twist angle, $\omega$, and the backbone contour length, $s$, per base-pair step. (c) Top view and side view of one base pair to illustrate the DNA diameter in our calculation and two dihedral angles, $\chi$ and $\delta$, mediating the twist-diameter coupling.

Our analysis of MD simulations, together with theoretical calculations, suggests a mechanism for the salt-induced twist change: the decrease of $c_{\text{salt}}$ induces the swelling of DNA diameter, which leads to the decrease of DNA twist through the twist-diameter coupling, as illustrated in Fig 3(a). In the following part, we will present the evidence for this mechanism and the calculations of the twist-diameter coupling constant.

Fig 4(a) shows a two-dimensional potential of mean force (PMF) as a function of the twist angle, $\omega$, and the diameter, $D$. This PMF, $P_{\text{sim}}$, was calculated from the simulation with 1 mol/L NaCl through

$$P_{\text{sim}}(\omega, D) = -(k_B T)\ln[\Omega(\omega, D)], \quad (1)$$

where $k_B$ is the Boltzmann constant, $T$ is the temperature, and $\Omega(\omega, D)$ is the relative density of DNA conformations for a given $\omega$ and $D$ from the simulation. Here, we define $D$ as the average distance between the two phosphate groups of one base pair. The position of the phosphate group is defined as the center of mass of one P and two O atoms.

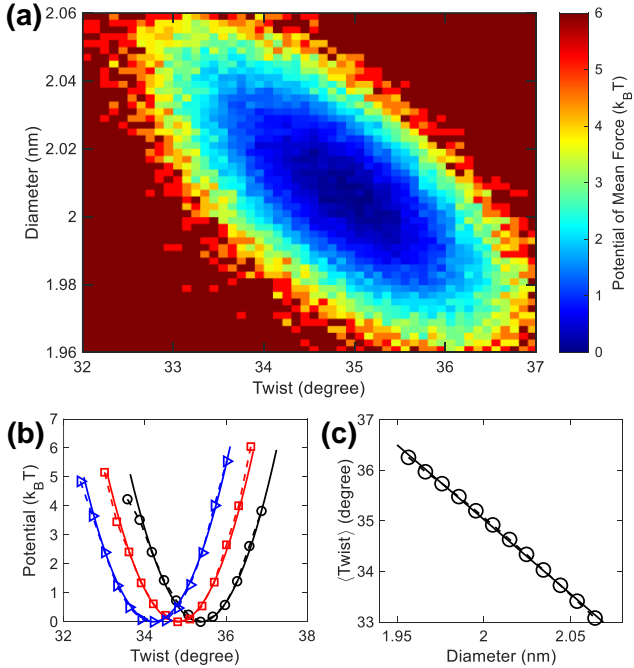

FIG.4 Free-energy analysis of the simulation result with 1 mol/L NaCl. (a) The two-dimensional potential of mean force (PMF) with respect to DNA twist angle and DNA diameter calculated from the simulation. (b) The one-dimensional PMF with respect to DNA twist angle for a narrow region of DNA diameter. The black, red, and blue symbols correspond to three regions of DNA diameter: $1.984 < D < 1.994$ nm, $2.004 < D < 2.014$ nm, $2.024 < D < 2.034$ nm, respectively. The solid lines are quadratic fits: $y = 1.7(x - 35.37)^2$, $y = 1.7(x - 34.81)^2$, and $y = 1.7(x - 34.21)^2$, respectively. (c) The average twist angle as a function of the diameter. The symbols are from our simulation, and the solid line is a linear fit: $y = -29.5x + 94.0$.

The PMF at 1 mol/L NaCl in Fig 4(a) has a global minimum at $\omega_0 \approx 34.82°$ and $D_0 \approx 2.008$ nm. The two-dimensional PMF exhibits a valley extending from the left-top to the right-bottom, indicating a negative coupling between $\omega$ and $D$. Similar to the analysis of twist-stretch coupling [1,15], we extracted the twist-diameter coupling constant by fitting the PMF using the following equation:

$$P_{\text{bp}} \approx k_\omega^{\text{bp}}(\Delta\omega)^2/2 + k_D^{\text{bp}}(\Delta D)^2/2 + k_{\omega D}^{\text{bp}}\Delta\omega\Delta D \quad (2)$$

and obtained the following coefficients for 1 mol/L NaCl:

$$k_\omega^{\text{bp}} \approx 0.18 \pm 0.02 \; k_B T/\text{deg}^2$$
$$k_D^{\text{bp}} \approx 263 \pm 39 \; k_B T/\text{nm}^2$$
$$k_{\omega D}^{\text{bp}} \approx 4.5 \pm 0.8 \; k_B T/(\text{deg} \cdot \text{nm}), \quad (3)$$

where $\Delta\omega \equiv \omega - \omega_0$, $\Delta D \equiv D - D_0$. The coefficients $k_\omega^{\text{bp}}$ and $k_D^{\text{bp}}$ characterize the rigidity of twist and diameter, respectively, while $k_{\omega D}^{\text{bp}}$ is the twist-diameter coupling constant. Note that these coefficients correspond to the relevant free energy per base pair as indicated by the superscript of "bp". We determined $k_\omega^{\text{bp}}$, $k_D^{\text{bp}}$, and $k_{\omega D}^{\text{bp}}$ by the fit to the PMF in Fig 4(a) using Eq (2). The uncertainties in Eq (3) correspond to 95% confidence interval during the fitting. We can convert $k_D^{\text{bp}} \approx 263 \; k_B T/\text{nm}^2$ to DNA Young's modulus of $3.2 \times 10^8$ Pa in agreement with experimental value of $3.46 \times 10^8$ Pa [31]. As indicated by Eq (2) and shown in Fig 4(b), for a given $D$, the PMF can be approximated by a harmonic potential, and the location of the potential minimum shifts toward a smaller $\omega$ with the increase of $D$. Fig 4(c) confirms that the dependence of average twist $\langle\omega\rangle$ on $D$ can be well captured by a straight line for a wide range of $D$.

Recall that the values of $\omega_0$, $D_0$, $k_\omega^{\text{bp}}$, $k_D^{\text{bp}}$, and $k_{\omega D}^{\text{bp}}$ in Eq (3) were calculated from the simulation with 1 mol/L NaCl. We also calculated $\omega_0$, $D_0$, $k_\omega^{\text{bp}}$, $k_D^{\text{bp}}$, and $k_{\omega D}^{\text{bp}}$ from the simulations with other $c_{\text{salt}}$ and the salt of KCl and RbCl (See Sec. S5). We find that $k_\omega^{\text{bp}}$, $k_D^{\text{bp}}$, and $k_{\omega D}^{\text{bp}}$ do not vary much for all three salt species of NaCl, KCl and RbCl and all $c_{\text{salt}}$ in our simulations from 0.05 to 1 mol/L, which agrees with our theoretical prediction (See Sec. S10 in the SI). In all of our simulations, $k_{\omega D}$ varies between 3.1 and 5.3 $k_B T/(\text{deg·nm})$. Based on the values of $k_\omega^{\text{bp}}$, $k_D^{\text{bp}}$, and $k_{\omega D}^{\text{bp}}$ in Eq (2), we can determine the twist rigidity for a relaxed DNA, by $\tilde{k}_\omega^{\text{bp}} = k_\omega^{\text{bp}} - (k_{\omega D}^{\text{bp}})^2/k_D^{\text{bp}} \approx 0.103 \; k_B T/\text{deg}^2$, which corresponds to a twist rigidity of 470 pN·nm$^2$ and agrees with previous experimental [32-35] and simulation [15,16] results (see Sec S9).

After quantifying the twist-diameter coupling by Eq (2), we proceed to the theoretical calculations of how the twist-diameter coupling leads to the salt-induced twist change. The basic idea of our calculation is as follows. The electrostatic repulsions between negatively charged P atoms on two DNA strands produce a force, $f_D$, to increase DNA diameter. The formula of $f_D$ has been derived by Manning using the helical distribution of charges [36-38]. It is expected that the decrease in $c_\text{salt}$ increases $f_D$ and enlarges $D$, which eventually leads to the decrease in $\omega$. To capture this salt-dependence twist change, we add a term $-f_D D$ into the PMF in Eq (2). Note that in the term of salt-dependence twist change, we need the change of $f_D$ upon $c_\text{salt}$ variation instead of the absolute value of $f_D$. Accordingly, we define the relative $\Delta f_D$ in line with the PMF in Eq (2):

$$\Delta f_D(c_\text{salt}) \approx f_D(c_\text{salt}) - f_D(1\,\text{mol/L}). \quad (4)$$

Then, we can write the PMF as

$$P(c_\text{salt}) \approx -\Delta f_D \Delta D + \frac{1}{2} k_\omega^\text{bp}(\Delta\omega)^2 + \frac{1}{2} k_D^\text{bp}(\Delta D)^2 + k_{\omega D}^\text{bp} \Delta\omega \Delta D, \quad (5)$$

The term of $-\Delta f_D \Delta D$ is an approximation based on the fact that for a given $c_\text{salt}$, $\Delta f_D$ remains almost unchanged within the small $D$ range we are dealing with. Minimizing $P(c_\text{salt})$ with respect to $\Delta\omega$ and $\Delta D$ yields

$$\Delta\omega(c_\text{salt}) = \frac{-k_{\omega D}^\text{bp}}{k_\omega^\text{bp} k_D^\text{bp} - \left(k_{\omega D}^\text{bp}\right)^2} \Delta f_D. \quad (6)$$

Substituting $\Delta f_D(c_\text{salt})$ calculated from the Manning formula (see Sec. S3 in the SI) into Eq (6), we obtained $\Delta\omega(c_\text{salt})$ as shown by the black line of Fig 2(a), which fairly agrees with the experimental and simulation results.

To validate the above theoretical calculations, we performed two additional sets of simulations. In one set of simulations, we manually added external forces (equivalent to $\Delta f_D$) on P atoms of DNA and then measured the change in DNA twist. We obtained results in excellent agreement with Eq (6) (see Fig S9). In the other set of simulations, we artificially modified the charge of each P atom from -1$e$ to -1.5$e$ or -0.5$e$. As expected, we observed a reduction of $\omega$ for the P charge of -1.5$e$ and a rise of $\omega$ for the P charge of -0.5$e$, compared with the case of the P charge of -1$e$. (See Fig S10). These results confirm that the salt-induced twist change is caused by the twist-diameter coupling and the variation of P-P electrostatic repulsions.

It is of interest to find out the molecular mechanism underlying the twist-diameter coupling. One likely mechanism is that the twist-diameter coupling is caused by the restraint of the contour length of DNA backbone per base-pair step, $s$, as illustrated in Fig 3(b). We can approximate $s \approx \sqrt{[(\omega D/2)^2 + h^2]}$, where $h$ is the rise along the helical axis per base-pair step. Due to the restraint of $s$, the increase of $D$ leads to the decrease of $\omega$. The stretch modulus of DNA backbone was estimated to be 965 pN by Gore et al [1]. Using this stretch modulus, we estimate the twist-diameter coupling constant as $k_{\omega D}^\text{bp} \approx 2.8\,k_B T/(\text{deg}\cdot\text{nm})$, which fairly agrees with the value in Eq (3) considering that we neglect many factors (see Sec S12). This proposed mechanism is consistent with results in previous studies [1,15,16], in particular, the excellent rod-wire model of DNA proposed by Gore et al [1].

We analyzed the structural basis of the twist-diameter coupling and found that the coupling is mediated through two dihedral angles in DNA backbone as illustrated in Fig 3(c). These two dihedral angles are commonly referred as $\chi$ and $\delta$. We found that both $\chi$ and $\delta$ are strongly and negatively correlated with the diameter, while strongly and positively correlated with the twist. The increase of $D$ under a lower salt concentration is realized through the decrease of $\chi$ and $\delta$, which simultaneously reduces $\omega$ (see Sec. S4).

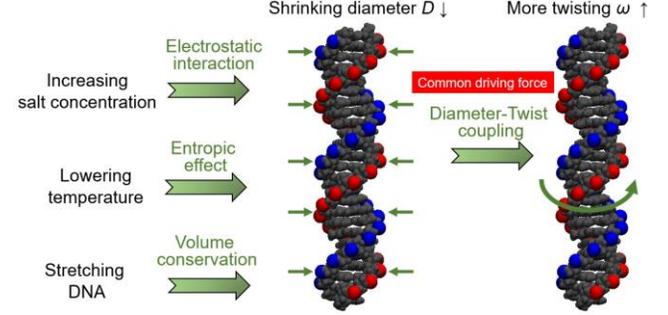

FIG.5 Twist-diameter coupling as a common driving force for stretch-, salt-, and temperature-induced DNA twist change.

Our analysis suggests that the twist-diameter coupling is prevalent in DNA deformations and is the common driving force for salt-, temperature-, and stretch-induced DNA twist changes (Fig 5). Lowering temperature increases DNA twist [4]. We propose a mechanism for this phenomenon: lowering temperature reduces DNA diameter through an entropic effect, which increases the twist through the coupling. Here, the entropic effect refers to the fact that DNA with a larger diameter has more conformational entropy, which is favored at higher temperature (see Fig S15). Using the twist-diameter coupling constant, we predicted $\Delta\omega_\text{bp}/\Delta T \approx -0.0102$ deg/K, which agrees with our and previous experimental results [4] (see Sec S11). Stretching DNA also increases DNA twist, which has been explained using a simple DNA model by Gore et al [1]. In this model, due to DNA volume conservation, stretching DNA leads to the shrinking of DNA diameter, which increases the twist through the coupling [1] (see Sec S12).

Varying $c_\text{salt}$ should affect many DNA structural parameters, not only DNA diameter, but also others, including the contour length $L$ and persistence length $L_p$ [6,7]. Our analysis suggests that the effect of salt on DNA twist is mainly mediated by diameter variation rather than variations of other structural parameters for the following reasons. First, experimental salt-induced twist change was quantitatively reproduced by salt-induced diameter variation and twist-diameter coupling. Second, adjusting diameter in simulations by external forces quantitatively reproduced twist change (Fig S9). Third, theoretical calculation of electrostatic interaction quantitatively reproduced salt-induced diameter variation. Fourth, twist-diameter coupling constant obtained from MD simulation fairly agrees with the one estimated the stretch modulus of DNA backbone, which suggests twist variation is directly linked to diameter variation rather than through a hidden structural parameter. Lastly, experimental temperature-induced twist change was quantitatively reproduced by diameter-dependent DNA conformational entropy and twist-diameter coupling (see Sec S11).

The central role of diameter in salt- and temperature-induced DNA deformations has physical origin. Diameter quantifies phosphate-phosphate distance, which adjusts charge interactions in salt effects, and quantifies inter-strand distance, which determines DNA conformational entropy.

In conclusion, we find that salt- and temperature-induced DNA twist changes are driven by twist-diameter coupling. For salt effects, this coupling takes effect with salt-induced diameter variation through electrostatic interactions. For temperature effects, this coupling takes effect with temperature-induced diameter variation through DNA conformational entropy. We determined twist-diameter coupling constant as approximately $4.5 \pm 0.8\ k_BT/(\text{deg}\cdot\text{nm})$ for one base pair, based on which we quantitatively experimental salt- and temperature-induced DNA twist changes. It is worth noting that although DNA diameter was not directly measured in experiments, extensive simulation and theoretical calculations seamlessly filled up all gaps that are needed to demonstrate twist-diameter coupling but unmeasurable in experiments (see details in Sec S13).

Salt-induced and temperature-induced DNA twist changes should have significant implications for relevant biological processes [19] and material applications of DNA origami [13,14]. Twist changes accumulate along DNA and can reach huge values. For 13.6 kbp DNA, the twist change can cause DNA rotation of several turns (Fig 1). In cells, the twist change may cause thousands to millions of rotations for mega to giga bp long.

We are grateful to the financial support from National Natural Science Foundation of China (No. 12074294 and 21973080) and the Research Grants Council of Hong Kong (Project No. 21302520).

#Chen Zhang and Fujia Tian contributed equally to this work.
Corresponding authors:
liangdai@cityu.edu.hk; zhxh@whu.edu.cn